\documentclass{article}

\usepackage[preprint]{neurips_2025}

\usepackage[utf8]{inputenc} % allow utf-8 input
\usepackage[T1]{fontenc}    % use 8-bit T1 fonts
\usepackage{hyperref}       % hyperlinks
\usepackage{url}            % simple URL typesetting
\usepackage{booktabs}       % professional-quality tables
\usepackage{amsfonts}       % blackboard math symbols
\usepackage{nicefrac}       % compact symbols for 1/2, etc.
\usepackage{microtype}      % microtypography
\usepackage{xcolor}         % colors

\usepackage{amsmath}
\usepackage{amssymb}
\usepackage{mathtools}
\usepackage{multirow}

\title{Revisiting Deep AC-OPF}

\author{
  Oluwatomisin I. Dada \\
  University of Cambridge \\
  \texttt{oid20@cam.ac.uk} \\
  \And
  Neil D. Lawrence \\
  University of Cambridge \\
  \texttt{ndl21@cam.ac.uk} \\
}

\begin{document}

\maketitle

\begin{abstract}
Recent work has proposed machine learning (ML) approaches as fast surrogates for solving AC optimal power flow (AC-OPF), with claims of significant speed-ups and high accuracy. In this paper, we revisit these claims through a systematic evaluation of ML models against a set of simple yet carefully designed linear baselines. We introduce \textbf{OPFormer-V}, a transformer-based model for predicting bus voltages, and compare it to both the state-of-the-art DeepOPF-V model and simple linear methods. Our findings reveal that, while OPFormer-V improves over DeepOPF-V, the relative gains of the ML approaches considered are less pronounced than expected. Simple linear baselines can achieve comparable performance. These results highlight the importance of including strong linear baselines in future evaluations.
\end{abstract}

\section{Introduction}
Electricity is an integral part of modern society; its reliable and efficient distribution is essential for sustaining many of the activities that define contemporary life. Amidst growing global concerns over climate change, there is an increasing desire to further electrify industries and incorporate more renewable energy sources - such as wind and solar - into our generation mix in an effort to reduce carbon emissions and mitigate the impacts of climate change. However, increased uncertainty due to variable renewable generation and highly dynamic loads means that network operators must solve AC optimal power flow (AC-OPF) problems more frequently. This need has sparked a lot of research on how best to apply machine learning (ML) approaches to this problem \citep{donti2021machine,donti2020enforcing,huang2021deepopf,donon2020neural,owerko2022unsupervised}.

These ML approaches can be broadly grouped into two (2) classes, \textit{direct} and \textit{hybrid} \citep{falconer2022leveraging}. \textit{Direct} approaches directly learn a mapping between the grid parameters and the OPF solution and are typically much faster than using a conventional solver or the \textit{hybrid} approach as they bypass the solver completely, however, they offer no feasibility guarantees \citep{falconer2020deep, 9053140, 9847037, donti2021dc3, liu2022topology, huang2021deepopf}. \textit{Hybrid} approaches predict information to help a conventional optimisation solver converge to a solution faster. \textit{Hybrid} approaches usually involve predicting primal and/or dual variables to warm start the optimiser or identifying non-binding constraints to formulate a reduced OPF problem which is passed to an optimiser/solver. This approach typical guarantees a feasible and optimal solution due to the optimiser/solver step \citep{robson2019learning, pham2022reduced, falconer2022leveraging}.

In this paper, we consider DeepOPF-V a state of the art \textit{direct} approach to solving AC-OPF developed by \citet{huang2021deepopf}. DeepOPF-V employs a fully connected model, but the novelty of this approach is primarily in the output targets it predicts. DeepOPF-V predicts the voltage magnitude ($V_{m}$) and angle ($V_{a}$) at each bus in the power grid and uses this to calculate the corresponding output of the generator ($S_{g}$) given the admittance ($Y_{bus}$) of the grid and the load ($S_{l}$). This approach allows the model to constrain the voltage magnitude and angle within the permissible range, and the remaining inequality and equality constraints are satisfied depending on the accuracy of the prediction. This approach is considered state-of-the-art based on the large speedups, high feasibility rate, and low optimality gap achieved by the model.

In this work, we contextualise the performance of DeepOPF-V through comparison against simpler benchmark models and introduce a transformer-based model, OPFormer-V, which predicts the same output targets as DeepOPF-V. Although this transformer-based model outperforms DeepOPF-V in all the datatsets considered, it performs worse than some simple benchmarks on the dataset OPF-Learn case 30 \citep{Joswig_Jones_2022}.

\begin{itemize}
    \item Implementation of a transformer-based model, OPFormer-V, which consistently outperformed the Fully Connected Network (FCNN) for all datasets considered.
    \item Comparison of a FCNN and a transformer-based model against relatively simple models and an observation of the surprisingly high performance achieved by simple linear models, including outperforming both neural network models on the dataset OPF-Learn case 30.
    \item Implementation of a linear OPF model that is based on a warm-start 1st-order Taylor approximation, with an error bound analysis for choice of reference point and performance comparison against DC-OPF amongst other methods.
\end{itemize}
\section{Background}
\subsection{Optimal power flow (OPF)}
OPF is a constrained optimisation problem that seeks to minimise the total cost of generation needed to satisfy demand while satisfying numerous equality and inequality constraints. A core component of OPF is the power flow equations, which describe the flow of power in the network. Consider a simplified power grid described by a graph $\mathcal{G}$ where the nodes in the graphs represent buses and the edges represent transmission lines connecting buses. This power grid also has an admittance matrix $\mathbf{Y}$, the off-diagonal elements of the matrix give the admittance between any pair of nodes in the graph $i,j$ and is zero when there is no transmission line connecting the nodes, while the diagonal elements of the matrix give the self-admittance of a node.

In a concise form, OPF can be expressed as shown in equation \ref{eq:opf4}, where $\mathbf{x}$ is the vector of grid parameters and $\mathbf{z}$ is the vector of optimization variables, $f(\mathbf{x}, \mathbf{z})$ is the objective function to minimize, subject to equality constraints $c_{j}^{E}(\mathbf{x}, \mathbf{z}) \in \mathcal{C}^{E}$ and inequality constraints $c_{k}^{I}(\mathbf{x}, \mathbf{z}) \in \mathcal{C}^{I}$,
\begin{equation} \label{eq:opf4}
    \begin{aligned}
        \min_{\mathbf{z}} f(\mathbf{x}, \mathbf{z}) \\
        \textrm{s.t.   } c_{j}^{E}(\mathbf{x}, \mathbf{z}) = 0 && j = 1 \dots m\\
        c_{k}^{I}(\mathbf{x}, \mathbf{z}) \geq 0 && k = 1 \dots n.
    \end{aligned}
\end{equation}
OPF can be considered an operator $\Phi (\cdot)$ that takes as input the load at each node $\mathbf{S_{l}}$, the admittance matrix of the grid $\mathbf{Y}$, the objective function $f(\cdot)$, the set of equality constraints $\mathcal{C}^{E}$ and inequality constraints $\mathcal{C}^{I}$ and produces as output the set points of the generator $\mathbf{S_{g}}$ (along with other variables).

There are numerous variations of OPF, with different variations having different additional constraints. In this work, we consider simple economic dispatch.

\subsubsection{Equality constraints}
The power flow equations that describe the active and reactive flow in an electrical power network are the set of equality constraints that must be satisfied in an AC-OPF problem. This can be formulated as shown in equation \ref{eq:opf1} as the power balance constraint, which simply states that at each node the difference between the generated and used power must be equal to the net power injected into the node,
\label{sssec:eq_const}
\begin{equation}
    \mathbf{S}_{inj} =\mathbf{S}_{g} - \mathbf{S}_{l} = \mathbf{V} \odot \mathbf{I}^{*} = \mathbf{V} \odot (\mathbf{Y}\mathbf{V})^{*},
    \label{eq:opf1}
\end{equation}
where $\mathbf{S}_{inj}$ is the difference between the complex power generated, $\mathbf{S}_{g}$, and the complex power consumed, $\mathbf{S}_{l}$, at each node. This difference is equal to the complex power that flows into the nodes from the connected nodes. $\mathbf{V}$ is a vector of complex voltage at each node, $\mathbf{I}$ is the net complex current injected at each node, $\odot$ is an Hadamard (element-wise) product, and $^{*}$ is the complex conjugate operator.

\subsubsection{Inequality constraints}
The inequality constraints that we consider include bounds on the voltage magnitudes at all nodes, real and reactive power generation, and an upper bound on the magnitude of complex power/current flowing through a transmission line.
\section{Related works}
As mentioned briefly in the introduction, ML approaches to OPF can be broadly grouped into two (2) classes, \textit{direct} and \textit{hybrid}. \textit{Direct} approaches directly learn a mapping between the grid parameters and the OPF solution while \textit{Hybrid} approaches predict information to help a conventional optimiser/solver converge to a solution faster.  In this section, we provide a brief overview of work done in both categories with more emphasis on direct approaches, and we also discuss linear power flow approximations.

\subsection{Direct approaches} 
\citet{9053140} used GNNs in a supervised manner to solve OPF problems directly predicting the active generation setpoints. Later in \citet{owerko2022unsupervised} they employed an unsupervised approach still using GNNs to solve the OPF problem. \citet{9847037} solve the OPF on the dual of the graph of the grid's topology, they predict current and power injections at each branch, by considering the line graph version of the grid, an approach they claim is more flexible to changes in the topology of the network. The work of \citet{falconer2020deep} compared different architectures and evaluated their ability to correctly predict OPF solutions for various grid topologies. Work by \citet{liu2022topology} employ GNNs as an adaptive OPF solver, in this work they also employ a feasibility regulariser which they term `physics-aware', it penalises violations of feasibility constraints and feasibility can be strictly enforced via projection. \citet{donti2021dc3} takes a more general approach to building a model for optimisations problems with feasibility guarantees via a differentiable procedure, which implicitly completes partial solutions to satisfy equality constraints and unrolls gradient-based corrections to satisfy inequality constraints. \citet{huang2021deepopf} introduced DeepOPF-V which predicts voltage magnitude and angles and uses those to calculate generator setpoints. \citet{9940481} and \citet{liang2023deepopfuunifieddeepneural} extend this work by employing the DeepOPF-V framework but training a single model across flexible topologies and various grids respectively.

\subsection{Hybrid approaches}
Hybrid approaches involve predicting primal and/or dual variables to warm start the optimiser or identifying non-binding constraints to formulate a reduced OPF problem which is passed to a solver. This approach typical guarantees a feasible and optimal solution because of the optimiser although it is also typically slower than direct approaches.  In \citet{robson2019learning} they learn an optimally reduced formulation of OPF through meta-optimisation, the model predicts a reduced OPF problem which is iteratively expanded until all biding constraints are considered and a meta-objective of reducing total computational time is included. \citet{pham2022reduced} they develop a GNN to reduce OPF and \citet{falconer2022leveraging} compares the performance of different architectures in predicting the binding status of lines. \citet{baker2022emulatingacopfsolvers} trains a DNN to emulate an iterative solver and, once close to convergence, passes the output to a power flow solver. \citet{piloto2024canos} trains a heterogeneous GNN to predict values which are then passed to a power flow solver.

\subsection{Linear power flow}
The primary source of non-convexity in AC-OPF is the power flow equation, there are numerous linear approximations of this equation that have been developed to make OPF convex, the most widely known version of this being DC-OPF. Numerous works \citep{6495738, 8119868, Yang2017ANN, 7782382, 7954975, 8468081} have focused on the development of different approaches to linearising power flow, considering different selections and transformations of the variables in the equation. \citet{6495738} propose a model that incorporates reactive power flows, \citet{8119868}'s approach involves a logarithmic transform of the voltage. \citet{8119868} propose using the square of the voltage instead of the voltage as an independent variable in the linear model. \citet{9916903} numerically compare using 7 different linearization approaches in multiple OPF problems and conclude that DC-OPF achieves the best overall performance.

\citet{9949682} consider an approach that uses the previous timestep AC-OPF solution as the reference point in a 1st order Taylor series instead of the flat start assumption typically used and found that this can reduce linearisation error when the load variations between adjacent time periods are not significant.

\section{Datasets}
In this work, we evaluated our methods on 3 datasets, 2 of which were self-generated and the other generated by \citet{Joswig_Jones_2022} as part of their OPF-Learn Dataset. All the datasets considered are synthetic.

In OPF research in the ML community, it is common to self-generate datasets. For this work the self-generated datasets are based on the IEEE case 30 and IEEE case 118 grids, having 30 and 118 buses/nodes respectively. The dataset were generated following the common approach of taking the nominal load case of the format and sampling random variations around that nominal loading scenario. We consider a $\pm50\%$ variation from the nominal load case at each node and perform latin hypercube sampling. This generates a loading scenario that is then solved using MATPOWER \citep{zimmerman1997matpower} and only scenarios that converge to a solution are included in the dataset. The 30 bus and 118 bus dataset contain 100k samples each.

The OPF-Learn dataset developed by \citet{Joswig_Jones_2022}, it does not employ the common approach to dataset generation, but is generated by trying to maximise the distinct number of active constraint sets included in the dataset, with the intention of generating a more comprehensive representation of the operating range of the grid. This dataset was retrieved from the NREL Data catalog \citep{joswig-jones2021opflearndata} from which we selected the 30 bus case, which contains 10k samples.

For all the datasets, we employ a 60/15/25 train/val/test data split.
\section{Methods}
We generally consider 7 different approaches to predicting voltage magnitude and angle, except for on the OPF-Learn Dataset for which we consider 5. The approaches considered are gridwise averaging, nodewise averaging, linear regression, DC-OPF, hot start linear power flow, DeepOPF-V and OPFormer-V.

\subsection{Baselines}
\subsubsection{Gridwise averaging}
This is the simplest heuristic, which simply takes the average across all nodes and across all samples for voltage magnitude and angle, respectively, in the training data and uses this as the prediction in the validation and testing data
\begin{equation}
    \hat{v}_{i} = \frac{1}{K}\sum^{K}_{j = 1}{\frac{1}{N}\sum^{N}_{i = 1}{v^{train}_{i,j}}}
    \label{eq1}.
\end{equation}
As shown in \ref{eq1} the predictor $\hat{v}$ is the same for all nodes $i$ and is the average $v^{train}$ value for all $N$ nodes over all $K$ training samples. This approach can be considered a data-driven flat start. 
This method is consistently the worst performing predictor of voltage angle as it assumes no power flow which is an assumption we know will be violated. However, it is a better predictor of voltage magnitude than DC-OPF which uses a fixed prediction of $1.0$ pu that is not data driven. 

\subsubsection{Nodewise averaging}
This heuristic computes an average across all samples in the training data for each node for voltage and magnitude
\begin{equation}
    \hat{v}_{i} = \frac{1}{K}\sum^{K}_{j = 1}{v^{train}_{i,j}}.
    \label{eq2}
\end{equation}
As shown in \ref{eq2} the predictor $\hat{v}$ for node $i$ is the average $v^{train}$ value for that node over all $K$ training samples. This method results in a fixed power flow between nodes that is the result of the difference between nodal voltage averages. Regression metrics show that this approach performs surprisingly well and is typically comparable to DeepOPF-V as shown in tables \ref{tab:ieee30_regression}, \ref{tab:ieee118_regression} and \ref{tab:opflearn_regression}.

\subsubsection{Linear regression}
This involves training 2$N$ linear models, where $N$ is the number of nodes/buses in the grid, to predict bus voltage magnitudes and angles. Each linear model takes the vector $\mathbf{S_{l}}$ as input and produces a single output and is Ordinary Least Square (OLS) regression. This approach does not predict a fixed output and consistently outperforms nodewise averaging on all datasets and is the best performing method on the OPF-Learn case 30 dataset as shown in tables \ref{tab:ieee30_regression}, \ref{tab:ieee118_regression} and \ref{tab:opflearn_regression}.

\subsection{Linear power flow}
In this work we consider two approaches that combine a linear approximation of the power flow with a conventional optimizer to solve an approximate version of the problem. The benefit of this approach is that it ensures generator outputs and nodal voltages are within proper bounds, however, as these values are no longer coupled via the actual power flow equations in order to test the feasiblity of the solution we select a subset of these variables and solve for the remaining variables using the power flow equations.

\subsubsection{DC-OPF}
Power flow is linearised in DC-OPF by assuming a voltage magnitude of $1.0$ pu at all nodes, using the small-angle approximation and the fact that the line inductance is typically much larger than the line resistance. These assumptions are equivalent to a 1st-order Taylor series with the flat start as the reference point and the additional assumption that the resistance of all lines is zero. This allows the power flow equations to be reduced to equation \ref{eq-dC-OPF:0}. Equation \ref{eq-dC-OPF:0} shows that in this approximation, the active power flow $p_{ij}$ from node $i$ to $j$ is determined solely by the angle difference between the nodes

\begin{equation}
    p_{ij} = -b_{ij}\left(\theta_{i} - \theta_{j}\right).
    \label{eq-dC-OPF:0}
\end{equation}

This heuristic was not readily available for the OPF-Learn Dataset and is not included as a benchmark for that dataset. From the regression metrics in tables \ref{tab:ieee30_regression} and \ref{tab:ieee118_regression} we can see that this method is a poor voltage magnitude predictor, it is a better voltage angle predictor but still only outperforms gridwise averaging for this target.

\subsubsection{Hot start}
Without the simplyfying assumptions made in DC-OPF, the linearised power flow equations are as shown in equations \ref{eq-linopf:0} - \ref{eq-linopf:3} where $\tilde{x}$ represents a variable at the reference point. In this work, we use CVXPY \citep{agrawal2018rewriting,diamond2016cvxpy} to solve the resulting convex optimisation problem
\begin{equation}
    \begin{split}
    p_{ij} =\tilde{p}_{ij}+ 2g_{ij}\tilde{v}_{i}\Delta_{v_{i}} + \sin\left(\tilde{\delta}_{ij}\right)\left[g_{ij}\tilde{v}_{i}\tilde{v}_{j}\Delta_{\delta_{ij}} - b_{ij}T_{v_{i},v_{j}}\right]\\ - 
    \cos\left(\tilde{\delta}_{ij}\right)\left[g_{ij}T_{v_{i},v_{j}} + b_{ij}\tilde{v}_{i}\tilde{v}_{j}\Delta_{\delta_{ij}}\right],
\end{split}
    \label{eq-linopf:0}
\end{equation}
\begin{equation}
    \begin{split}
        q_{ij} =\tilde{q}_{ij}  -2b_{ij}\tilde{v}_{i}\Delta_{v_{i}} - \sin\left(\tilde{\delta}_{ij}\right)\left[b_{ij}\tilde{v}_{i}\tilde{v}_{j}\Delta_{\delta_{ij}} + g_{ij}T_{v_{i},v_{j}}\right]\\ + 
    \cos\left(\tilde{\delta}_{ij}\right)\left[b_{ij}T_{v_{i},v_{j}} - g_{ij}\tilde{v}_{i}\tilde{v}_{j}\Delta_{\delta_{ij}}\right],
    \end{split}
    \label{eq-linopf:1}
\end{equation}
\begin{equation}
        T_{v_{i},v_{j}} = \tilde{v}_{i}\Delta_{v_{j}} + \tilde{v}_{j}\Delta_{v_{i}},
    \label{eq-linopf:2}
\end{equation}
\begin{equation}
        \Delta_{x_{k}} = x_{k} - \tilde{x}_{k}.
    \label{eq-linopf:3}
\end{equation}
This approach models both active and reactive power flow and implicitly linearises power loss around the reference point, as shown in equation \ref{eq-linopf:4}
\begin{equation}
        p_{ij}^{loss} = p_{ij} + p_{ji} = \tilde{p}_{ij}^{loss}+
        2g_{ij}\left[    \tilde{v}_{i}\tilde{v}_{j}\sin\left(\tilde{\delta}_{ij}\right)\Delta_{\delta_{ij}} +
        \left[1-\cos\left(\tilde{\delta}_{ij}\right)\right]T_{v_{i},v_{j}}
        \right].
    \label{eq-linopf:4}
\end{equation}
The reference point chosen is the nodewise voltage average which minimises the expected MAE on an upper bound of the truncation error ($R(\zeta)$) as shown in equation \ref{eq-linopf:6} for active power flow. This bound shows the MAE as a function of the 2nd-order moments of the variables $v_{i}$, $v_{j}$, $\delta_{ij}$ centred at $\tilde{v}_{i}$, $\tilde{v}_{j}$, $\tilde{\delta}_{ij}$. The 2nd-order moment of a random variable is minimised when centred at the mean of that variable
\begin{equation}
    \begin{split}
        R\left(\zeta\right) =  
        g_{ij}
    \Delta_{v_{i}}^{2} + 
    |y_{ij}|\left[
    \frac{\zeta_{{v}_{i}}\zeta_{{v}_{j}}}{2}\cos\left(\zeta_{{\delta}_{ij}} - \angle _{y_{ij}} \right)\Delta_{\delta_{ij}}^{2} -
    \Lambda\sin\left(\zeta_{{\delta}_{ij}} - \angle _{y_{ij}} \right)
    \right]
     \\
    \Lambda = \zeta_{{v}_{i}}\Delta_{v_{j}}\Delta_{\delta_{ij}}  + \Delta_{v_{i}}\zeta_{{v}_{j}}\Delta_{\delta_{ij}},
    \end{split}
    \label{eq-linopf:5}
\end{equation}
\begin{equation}
    \begin{split}
        \mathbb{E}\left[|R\left(\zeta\right)|\right] \leq  
        \left|g_{ij}\right|\mathbb{E}\left[\Delta_{v_{i}}^{2}\right] + 
    |y_{ij}|\left[
    \frac{v_{ub}^{2}}{2}\mathbb{E}\left[\Delta_{\delta_{ij}}^{2}\right] +
    \mathbb{E}\left[\left|\Lambda\right|\right]_{ub}\right]
    \\
    \mathbb{E}\left[\left|\Lambda\right|\right]_{ub} = v_{ub}\left[
    \left(\mathbb{E}\left[\Delta_{v_{j}}^{2}\right]\mathbb{E}\left[\Delta_{\delta_{ij}}^{2}\right]\right)^{\frac{1}{2}} +
    \left(\mathbb{E}\left[\Delta_{v_{i}}^{2}\right]\mathbb{E}\left[\Delta_{\delta_{ij}}^{2}\right]\right)^{\frac{1}{2}}
    \right].
    \end{split}
    \label{eq-linopf:6}
\end{equation}

For a data-efficient approach, we can approximate the average nodal voltages as the nodal voltages of the average loading scenario. The error in this approximation is small assuming that the mapping from load to voltage has a bounded second derivative and that the variance in loading scenarios is small; numerically, we observed that this was the case in our datasets.

\subsection{ML approaches}
\subsection{DeepOPF-V}
DeepOPF-V is an FCNN model trained with an L2 loss that jointly predicts the voltage angle and magnitude at all nodes in the grid. The original paper \citep{huang2021deepopf} also includes a post-processing step to help reduce generator limit violations; however, the effect of this step is minimal from the reported results. In comparing methods, we do not consider this additional post-processing step. This method is the foundation for state-of-the-art ML approaches to AC-OPF; however, when we contextualize it against much simpler models, we see that it performs similarly to nodewise averaging in predicting output targets as shown in tables \ref{tab:ieee30_regression}, \ref{tab:ieee118_regression} and \ref{tab:opflearn_regression}.

\subsection{OPFormer-V}
In this paper we introduce OPFormer-V which is an attention-based approach to solving the OPF problem. Grids with $N$ buses are inputs into a transformer encoder as a sequence with $N$ tokens, each token containing the relevant information at an individual node.  The encoder output sequence is then concatenated, passed to a simple feedforward network, and used to jointly predict voltage angles and magnitudes at all buses.

Apart from the load ($p_{l,i}$, $q_{l,i}$) at a node, this architecture also supports passing information on generator limits ($p^{max}_{g,i}$, $p^{min}_{g,i}$, $q^{max}_{g,i}$, $q^{min}_{g,i}$), generator costs $(c1, c2)$ and shunt impedances ($bs_{i}$, $gs_{i}$) at a node, which are features that vary between nodes but not between samples and as such are not included in DeepOPF-V which concatenates all nodal information into a single input vector.
\section{Evaluation}
In the following section, we evaluate the various methods considered in terms of regression metrics and power metrics. Regression metrics assess how accurately a method predicts voltage magnitude and angle and the power metrics assess how well the resulting generation using the predicted voltage aligns with AC-OPF and respects constraints. The results for DeepOPF-V and OPFormer-V are based on 3 runs with different seeds and are evaluated without the post-processing step included by \citet{huang2021deepopf}.

\subsection{Regression metrics}
Tables \ref{tab:ieee30_regression}, \ref{tab:ieee118_regression}, and \ref{tab:opflearn_regression} present the regression metrics for the IEEE-case30, IEEE-case118, and OPF-Learn case 30 datasets, respectively. In these tables, the FVU for the \textbf{Grid Average} method is consistently $1.000$. This is expected, as the Grid Average method uses the mean of the training data as the predictor, confirming that there is no shift in the mean of the distribution from the training to the test set. Additionally, the MSEs for this method provide the variance of the voltage magnitude and angle. Tables \ref{tab:ieee30_regression} and \ref{tab:ieee118_regression} show that all methods, except for \textbf{Grid Average}, outperform \textbf{DC-OPF} in predicting voltage magnitude and angle, with the \textbf{OPFormer-V} variation emerging as the best-performing method according to these metrics. Specifically, in table \ref{tab:ieee30_power}, \textbf{OPFormer-V, feats 8} outperforms \textbf{OPFormer-V, feats 2} on both metrics. In contrast, in table \ref{tab:ieee118_power}, the \textbf{feats 2} variation performs better in predicting the voltage angle. However, since the model is trained to jointly predict both voltage magnitude and angle, and given the relative difference in magnitude between these variables, \textbf{feats 8} achieves a lower loss value in both cases. The additional information provided in \textbf{feats 8} results in a slight decrease in loss compared to \textbf{feats 2}, although the joint prediction task may lead to trade-offs between the two variables. \textbf{DC-OPF} proves to be a better predictor for voltage angle than \textbf{Grid Average}. However, because it approximates the voltage magnitude as a constant $1.0$ pu, it performs worse than the mean for voltage magnitude (with mean voltage magnitudes of $1.0405$ and $1.0243$ for IEEEcase30 and IEEEcase118, respectively), leading to an increase in FVU. The tables also reveal that simple linear methods, such as \textbf{Node Average} and \textbf{Linear Regression}, perform surprisingly well. In particular, in table \ref{tab:ieee118_regression}, these methods narrow the gap with neural network methods, with \textbf{Linear Regression} achieving a lower FVU than DeepOPF-V.
% \usepackage{booktabs}
% \usepackage{multirow}

\begin{table}[!htbp]
\caption{A table showing the MSE and FVU in predicting the voltage angle ($V_{a}$) in rad and voltage magnitude ($V_{m}$) in pu of the 7 different methods considered for the test split on the self-generated dataset on the IEEE case 30 grid. There are 2 variations of the OPFormer-V shown, \textbf{feats-2} takes a 2 dimensional vector of load $(p_{l,i}, q_{l,i})$ as input while \textbf{feats-8} takes an eight dimensional vector of load, shunt susceptance and generator information $(p_{l,i}, q_{l,i}, bs_{i}, p^{max}_{g,i}, q^{max}_{g,i}, q^{min}_{g,i}, c1, c2)$. For NN methods we report the mean and standard deviation over 3 runs.}
\label{tab:ieee30_regression}
\centering
\begin{tabular}{lrr|rr}
\hline
\multicolumn{1}{c}{\multirow{2}{*}{\textbf{Method}}}                                                         & \multicolumn{2}{c|}{\textbf{$V_{a}$}}                                & \multicolumn{2}{c}{\textbf{$V_{m}$}}                                \\ \cline{2-5} 
\multicolumn{1}{c}{}                                                                                         & \multicolumn{1}{c}{\textbf{MSE}} & \multicolumn{1}{c|}{\textbf{FVU}} & \multicolumn{1}{c}{\textbf{MSE}} & \multicolumn{1}{c}{\textbf{FVU}} \\ \hline
\multicolumn{1}{l|}{\multirow{2}{*}{\textbf{DeepOPF-V}}}                                                     & $8.782 \times 10^{-6}$                       & $2.280 \times 10^{-3}$                        & $3.272 \times 10^{-6}$                       & $9.169 \times 10^{-3}$                       \\
\multicolumn{1}{l|}{}                                                                                        & $\pm 5.3 \times 10^{-7}$                     & $\pm 1.4 \times 10^{-4}$                      & $\pm 1.2 \times 10^{-7}$                     & $\pm 3.3 \times 10^{-4}$                     \\
\multicolumn{1}{l|}{\multirow{2}{*}{\textbf{\begin{tabular}[c]{@{}l@{}}OPFormer-V,\\ feats 2\end{tabular}}}} & $2.402 \times 10^{-7}$                       & $6.235 \times 10^{-5}$                        & $3.976 \times 10^{-8}$                       & $1.112 \times 10^{-4}$                       \\
\multicolumn{1}{l|}{}                                                                                        & $\pm 1.6 \times 10^{-7}$                     & $\pm 4.1 \times 10^{-5}$                      & $\pm 1.2 \times 10^{-8}$                     & $\pm 3.5 \times 10^{-5}$                     \\
\multicolumn{1}{l|}{\multirow{2}{*}{\textbf{\begin{tabular}[c]{@{}l@{}}OPFormer-V,\\ feats 8\end{tabular}}}} & $\mathbf{2.089 \times 10^{-7}}$              & $\mathbf{5.421 \times 10^{-5}} $              & $\mathbf{3.123 \times 10^{-8}}$              & $\mathbf{8.731 \times 10^{-5}}$              \\
\multicolumn{1}{l|}{}                                                                                        & $\pm 8.4 \times 10^{-8}$                     & $\pm 2.2 \times 10^{-5}$                      & $\pm 4.3 \times 10^{-9}$                     & $\pm 1.2 \times 10^{-5}$                     \\
\multicolumn{1}{l|}{\textbf{Grid Avg.}}                                                                      & $3.853 \times 10^{-3}$                       & $1.000\times 10^{-0}$                        & $3.577 \times 10^{-4}$                       & $1.000 \times 10^{-0}$                       \\
\multicolumn{1}{l|}{\textbf{Node Avg.}}                                                                      & $9.804 \times 10^{-5}$                       & $2.545 \times 10^{-2}$                        & $2.765 \times 10^{-5}$                       & $7.731 \times 10^{-2}$                       \\
\multicolumn{1}{l|}{\textbf{DC-OPF}}                                                                          & $1.932 \times 10^{-3}$                       & $5.015 \times 10^{-1}$                        & $9.493 \times 10^{-4}$                       & $2.654 \times 10^{-0}$                       \\
\multicolumn{1}{l|}{\textbf{Hot-Start PF}}                                                                          & $1.967 \times 10^{-5}$                       & $5.105 \times 10^{-3}$                        & $4.480 \times 10^{-5}$                       & $1.252 \times 10^{-1}$                       \\
\multicolumn{1}{l|}{\textbf{Linear}}                                                                         & $6.272 \times 10^{-6}$                       & $1.628 \times 10^{-3}$                        & $4.484 \times 10^{-7}$                       & $1.254 \times 10^{-3}$                       \\ \hline
%\multicolumn{1}{l|}{\textbf{GP}}                                                                             & $9.472 \times 10^{-7}$                       & $2.459 \times 10^{-4}$                        & $7.734 \times 10^{-8}$                       & $2.162 \times 10^{-4}$                       \\ \hline
\end{tabular}
\end{table}

\begin{table}[!htbp]
\caption{A table showing the MSE and FVU in predicting the voltage angle ($V_{a}$) in rad and voltage magnitude ($V_{m}$) in pu of the 7 different methods considered for the test split on the self-generated dataset on the IEEE case 118 grid. There are 2 variations of the OPFormer-V shown, \textbf{feats-2} takes a 2 dimensional vector of load $(p_{l,i}, q_{l,i})$ as input while \textbf{feats-8} takes an eight dimensional vector of load, shunt susceptance and generator information $(p_{l,i}, q_{l,i}, bs_{i}, p^{max}_{g,i}, q^{max}_{g,i}, q^{min}_{g,i}, c1, c2)$.  For NN methods we report the mean and standard deviation over 3 runs.}
\label{tab:ieee118_regression}
\centering
\begin{tabular}{lrr|rr}
\hline
\multicolumn{1}{c}{\multirow{2}{*}{\textbf{Method}}}                                                         & \multicolumn{2}{c|}{\textbf{$V_{a}$}}                                & \multicolumn{2}{c}{\textbf{$V_{m}$}}                                \\ \cline{2-5} 
\multicolumn{1}{c}{}                                                                                         & \multicolumn{1}{c}{\textbf{MSE}} & \multicolumn{1}{c|}{\textbf{FVU}} & \multicolumn{1}{c}{\textbf{MSE}} & \multicolumn{1}{c}{\textbf{FVU}} \\ \hline
\multicolumn{1}{l|}{\multirow{2}{*}{\textbf{DeepOPF-V}}}                                                     & $7.351\times 10 ^{-5}$                       & $1.090\times 10 ^{-2}$                        & $6.458\times 10 ^{-6}$                       & $4.129\times 10 ^{-2}$                       \\
\multicolumn{1}{l|}{}                                                                                        & $\pm4.4\times 10 ^{-6}$                      & $\pm6.5\times 10 ^{-4}$                       & $\pm9.5\times 10 ^{-8}$                      & $\pm6.1\times 10 ^{-4}$                      \\
\multicolumn{1}{l|}{\multirow{2}{*}{\textbf{\begin{tabular}[c]{@{}l@{}}OPFormer-V,\\ feats 2\end{tabular}}}} & $\mathbf{1.921\times 10 ^{-6}}$                       & $\mathbf{2.849\times 10 ^{-4}}$                        & $8.545\times 10 ^{-8}$                       & $5.464\times 10 ^{-4}$                       \\
\multicolumn{1}{l|}{}                                                                                        & $\pm 1.3\times 10 ^{-7}$                     & $\pm 1.9\times 10 ^{-5}$                      & $\pm 8.1\times 10 ^{-9}$                     & $\pm 5.2\times 10 ^{-5}$                     \\
\multicolumn{1}{l|}{\multirow{2}{*}{\textbf{\begin{tabular}[c]{@{}l@{}}OPFormer-V,\\ feats 8\end{tabular}}}} & $2.708\times 10 ^{-6}$              & $4.016\times 10 ^{-4}$               & $\mathbf{6.703\times 10 ^{-8}}$              & $\mathbf{4.286\times 10 ^{-4}}$              \\
\multicolumn{1}{l|}{}                                                                                        & $\pm3.2\times 10 ^{-7}$                      & $\pm4.7\times 10 ^{-5}$                       & $\pm5.4\times 10 ^{-9}$                      & $\pm3.4\times 10 ^{-5}$                      \\
\multicolumn{1}{l|}{\textbf{Grid Avg.}}                                                                      & $6.743\times 10 ^{-3}$                       & $1.000\times 10^{-0}$                        & $1.564\times 10 ^{-4}$                       & $1.000\times 10^{-0}$                       \\
\multicolumn{1}{l|}{\textbf{Node Avg.}}                                                                      & $4.447\times 10 ^{-4}$                       & $6.595\times 10 ^{-2}$                        & $7.528\times 10 ^{-6}$                       & $4.814\times 10 ^{-2}$                       \\
\multicolumn{1}{l|}{\textbf{DC-OPF}}                                                                          & $4.575\times 10 ^{-3}$                       & $6.785\times 10 ^{-1}$                        & $1.795\times 10 ^{-3}$                       & $1.148\times 10^{+1}$                       \\
\multicolumn{1}{l|}{\textbf{Hot-Start PF}}                                                                          & $1.174\times 10 ^{-3}$                       & $1.741\times 10 ^{-1}$                        & $2.519\times 10 ^{-3}$                       & $1.611\times 10^{+1}$                       \\
\multicolumn{1}{l|}{\textbf{Linear}}                                                                         & $2.986\times 10 ^{-6}$                       & $4.428\times 10 ^{-4}$                        & $1.188\times 10 ^{-7}$                       & $7.596\times 10 ^{-4}$                       \\ \hline
%\multicolumn{1}{l|}{\textbf{GP}}                                                                             & $8.929\times 10 ^{-6$                       & $1.324\times 10 ^{-3$                        & $3.938\times 10 ^{-7$                       & $2.518\times 10 ^{-3$                       \\ \hline
\end{tabular}
\end{table}

Table \ref{tab:opflearn_regression} shows that, in the OPF-Learn dataset, there is a decrease in the voltage angle variance, an increase in the voltage magnitude variance and a general increase in FVU. While \textbf{DC-OPF} was not considered for this dataset, most methods performed generally worse in terms of FVU, with the exception of \textbf{Grid Average}, which performed the same. Interestingly, \textbf{Linear Regression} emerged as the best performing method on this dataset, although this may be partly due to the small size of the dataset. For this table, the \textbf{OPFormer-V, feats 8} variation and \textbf{DC-OPF} were not considered, as the additional information required was not available in this external dataset.

\begin{table}[!htbp]
\caption{A table showing the MSE and FVU in predicting the voltage angle ($V_{a}$) in rad and voltage magnitude ($V_{m}$) in pu for the different methods considered on the test split on the OPF-Learn case 30 dataset. The OPFormer-V variation considered \textbf{feats-2} takes a 2 dimensional vector of load $(p_{l,i}, q_{l,i})$ as input. For NN methods we report the mean and standard deviation over 3 runs.}
\centering
\label{tab:opflearn_regression}
\begin{tabular}{lrr|rr}
\hline
\multicolumn{1}{c}{\multirow{2}{*}{\textbf{Method}}}                                                         & \multicolumn{2}{c|}{\textbf{$V_{a}$}}                                & \multicolumn{2}{c}{\textbf{$V_{m}$}}                                \\ \cline{2-5} 
\multicolumn{1}{c}{}                                                                                         & \multicolumn{1}{c}{\textbf{MSE}} & \multicolumn{1}{c|}{\textbf{FVU}} & \multicolumn{1}{c}{\textbf{MSE}} & \multicolumn{1}{c}{\textbf{FVU}} \\ \hline
\multirow{2}{*}{\textbf{DeepOPF-V}}&  $4.334\times 10 ^{-7}$                      & $1.393\times 10 ^{-1}$                       & $2.396\times 10 ^{-4}$                      & $2.750\times 10 ^{-1}$                      \\
                                    & $\pm 5.8\times 10 ^{-12}$                    & $\pm 1.9\times 10 ^{-6}$                     & $\pm 1.3\times 10 ^{-9}$                    & $\pm 1.5\times 10 ^{-6}$                    \\
\multirow{2}{*}{\textbf{\begin{tabular}[l]{@{}l@{}}OPFormer-V\\ (feats 2)\end{tabular}}}     & $3.458\times 10 ^{-7}$                      & $1.112\times 10 ^{-1}$                       & $2.205\times 10 ^{-4}$                      & $2.530\times 10 ^{-1}$                      \\
                                                                                          & $\pm 7.6\times 10 ^{-8}$                    & $\pm 2.4\times 10 ^{-2}$                     & $\pm 1.7\times 10 ^{-5}$                    & $\pm 2.0\times 10 ^{-2}$                    \\
\textbf{Grid Avg.}                                                                                  & $3.111\times 10 ^{-6}$                      & $1.000 \times 10^{0}$                       & $8.715\times 10 ^{-4}$                      & $1.000 \times 10^{0}$                      \\
\textbf{Node Avg.}                                                                                  & $4.334\times 10 ^{-7}$                      & $1.393\times 10 ^{-1}$                       & $2.396\times 10 ^{-4}$                      & $2.749\times 10 ^{-1}$                      \\
\textbf{Linear}                                                                                     & $\mathbf{7.404\times 10 ^{-9}}$              & $\mathbf{2.380\times 10 ^{-3}}$               & $\mathbf{3.475\times 10 ^{-5}}$              & $\mathbf{3.987\times 10 ^{-2}}$             \\
%\textbf{GP}                                                                                         & $7.555\times 10 ^{-9}$                       & $2.428\times 10 ^{-3}$                        & $3.532\times 10 ^{-5}$                       & $4.053\times 10 ^{-2}$                       \\ 
\hline
\end{tabular}
\end{table}

Overall, these results underscore the potential of transformer-based models like \textbf{OPFormer-V} for improving voltage prediction accuracy in AC-OPF problems, while also highlighting the unexpected robustness of simpler linear methods. This suggests that further exploration into the balance between model complexity and performance could be valuable, particularly in different dataset scenarios and for enhancing model generalisability.

\subsection{Power metrics}
In this section, we compare the performance of \textbf{DeepOPF-V} and \textbf{OPFormer-V} in terms of the optimality gap, generation limit violation rate, and error in effective load determined using the predicted voltage. As shown in tables \ref{tab:ieee30_power} and \ref{tab:ieee118_power}, \textbf{OPFormer-V} generally outperforms \textbf{DeepOPF-V} on both datasets. Regarding the optimality gap, when considering only the relative difference, both methods exhibit a difference of less than 1\%, consistent with the results presented in the original paper \citep{huang2021deepopf}. However, when we examine the absolute relative difference, we find that only \textbf{OPFormer-V} maintains a difference below 1\%. In terms of the generation limit violation rate, we report higher rates compared to the original paper, primarily due to the greater variation in load ($\pm0.5$ compared to $\pm0.1$). Although \textbf{OPFormer-V} has a slight advantage in violation rate, the comparable rates between the two methods suggest that improvements in voltage prediction at this error level do not translate linearly to reductions in violation rates. This finding also underscores the need for approaches that can account for both generator and voltage limits. Lastly, considering the relative error in the effective load, particularly when examining the aggregate and nonzero loads, \textbf{OPFormer-V} demonstrates a lower error.

\begin{table}[!htbp]
\caption{A table comparing the quality of the OPF solutions from the predictions of DeepOPF-V and OPFormer-V (feats-8) on the test split on the IEEE case30 datasets. Predictions are assessed on the average relative gap from optimality based on groundtruth, the rate of violation of generation limits, the average relative difference between load in the ground truth and effective load derived using predicted voltage for both a grid aggregation and at a nodal level for $\neq 0$ loads.}
\label{tab:ieee30_power}
\centering
\begin{tabular}{lrl|rl}
\hline
\multicolumn{1}{c}{\textbf{IEEE case30}}                                   & \multicolumn{2}{c|}{\textbf{DeepOPF-V}} & \multicolumn{2}{c}{\textbf{OPFormer-V}} \\ \hline
\multicolumn{1}{l|}{\textbf{Rel. Opt. Diff. (\%)}}                    & -0.025           & $\pm 0.082$          & \textbf{0.087}            & $\pm 0.095$          \\
\multicolumn{1}{l|}{\textbf{Abs. Rel. Opt. Diff. (\%)}}               & 2.427            & $\pm 0.023$          & \textbf{0.150}            & $\pm 0.050$          \\
\multicolumn{1}{l|}{\textbf{$\mathbf{P_{g}}$ Violation Rate (\%)}}         & 10.984           & $\pm 0.255$          & \textbf{10.488}           & $\pm 0.639$          \\
\multicolumn{1}{l|}{\textbf{$\mathbf{Q_{g}}$ Violation Rate (\%)}}         & \textbf{14.932}           & $\pm 0.418$          & 16.629           & $\pm 1.392$          \\
\multicolumn{1}{l|}{\textbf{Abs. Rel. Tot. $\mathbf{P_{d}}$ err. (\%)}}    & 1.876            & $\pm 0.019$          & \textbf{0.116}            & $\pm 0.038$          \\
\multicolumn{1}{l|}{\textbf{Abs. Rel. Tot. $\mathbf{Q_{d}}$ err. (\%)}}    & 2.151            & $\pm 0.029$          & \textbf{0.144}            & $\pm 0.014$          \\
\multicolumn{1}{l|}{\textbf{Abs. Rel. $\mathbf{P^{\neq 0}_{d}}$ err. (\%)}} & 22.270           & $\pm 0.138$          & \textbf{2.251}            & $\pm 0.208$          \\
\multicolumn{1}{l|}{\textbf{Abs. Rel. $\mathbf{Q^{\neq 0}_{d}}$ err. (\%)}} & 23.627           & $\pm 0.052$          & \textbf{6.657}            & $\pm 0.657$          \\ \hline
\end{tabular}
\end{table}

In general, these results highlight the superior performance of \textbf{OPFormer-V} across multiple metrics, while also revealing areas where further improvements and more comprehensive approaches may be necessary to fully optimize power flow solutions.

\begin{table}[!htbp]
\caption{A table comparing the quality of the OPF solutions from the predictions of DeepOPF-V and OPFormer-V (feats-8) on the test split on the IEEE case118 datasets. Predictions are assessed on the average relative gap from optimality based on groundtruth, the rate of violation of generation limits, the average relative difference between load in the ground truth and effective load derived using predicted voltage for both a grid aggregation and at a nodal level for $\neq 0$ loads.}
\label{tab:ieee118_power}
\centering
\begin{tabular}{lrl|rl}
\hline
\multicolumn{1}{c}{\textbf{IEEE case118}}                                       & \multicolumn{2}{c|}{\textbf{DeepOPF-V}} & \multicolumn{2}{c}{\textbf{OPFormer-V}} \\ \hline
\multicolumn{1}{l|}{\textbf{Rel. Opt. Diff. (\%)}}                    & -0.618           & $\pm 0.012$          & \textbf{-0.153}           & $\pm 0.053$          \\
\multicolumn{1}{l|}{\textbf{Abs. Rel. Opt. Diff. (\%)}}               & 1.713            & $\pm 0.018$          & \textbf{0.323}            & $\pm 0.045$          \\
\multicolumn{1}{l|}{\textbf{$\mathbf{P_{g}}$ Violation Rate (\%)}}         & 21.799           & $\pm 0.044$          & \textbf{16.468}           & $\pm 0.306$          \\
\multicolumn{1}{l|}{\textbf{$\mathbf{Q_{g}}$ Violation Rate (\%)}}         & 12.605           & $\pm 0.114$          & \textbf{11.771}           & $\pm 0.742$          \\
\multicolumn{1}{l|}{\textbf{Abs. Rel. Tot. $\mathbf{P_{d}}$ err. (\%)}}    & 1.242            & $\pm 0.012$          & \textbf{0.245}            & $\pm 0.035$          \\
\multicolumn{1}{l|}{\textbf{Abs. Rel. Tot. $\mathbf{Q_{d}}$ err. (\%)}}    & 1.472            & $\pm 0.006$          & \textbf{0.241}            & $\pm 0.011$          \\
\multicolumn{1}{l|}{\textbf{Abs. Rel. $\mathbf{P^{\neq 0}_{d}}$ err. (\%)}} & 16.242           & $\pm 0.140$          & \textbf{4.053}            & $\pm 0.182$          \\
\multicolumn{1}{l|}{\textbf{Abs. Rel. $\mathbf{Q^{\neq 0}_{d}}$ err. (\%)}} & 17.648           & $\pm 0.252$          & \textbf{4.934}            & $\pm 0.112$          \\ \hline
\end{tabular}
\end{table}
\section{Conclusion}
We introduced \textbf{OPFormer-V}, a transformer-based model for predicting voltages to solve the AC-OPF problem. We evaluated \textbf{OPFormer-V}  against \textbf{DeepOPF-V} in three datasets and demonstrated superior performance in both regression and power metrics. We benchmarked both models against simpler linear models, demonstrating that simpler models can achieve comparable regression performance despite the non-linear nature of AC-OPF. 

The linear models \textbf{Node Average} and \textbf{Linear Regression} outperformed \textbf{DC-OPF} on the self-generated IEEEcase30 and IEEEcase118 datasets, despite \textbf{DC-OPF} being an optimization problem based on a linearized power flow. This is likely due to two factors. Firstly, the data generation method of varying around a nominal load case appears to result in relatively small variances in the magnitude and angle of the voltage at each node. Although this is a widely adopted data generation process and reflects conditions similar to those grid operators face, even a restricted fast OPF solver can be of great utility to grid operators. However, these results suggest that the current stage of ML solvers offers only marginal improvements over benchmark linear methods. Secondly, the approximations used to linearise power flow in \textbf{DC-OPF} involve additional approximations in a first-order Taylor expansion. These approximations can reduce the accuracy of the prediction \citep{6495738, 9916903}, we observed a reduction in the voltage angle error and in the optimality gap when we used the hot start linear power flow approach.

When we consider power metrics, we find that even models with good regression metrics can still have significant generation limit violations, highlighting the need for models that respect all constraints.

\bibliography{revisit_opf}
\bibliographystyle{plainnat}

\newpage
\appendix
\section{Appendix}
\subsection{Hot-start linear power flow}
\subsubsection{MAE upper bound}
For a first-order Taylor series approximation of a function the error in approximation is given by the remainder term $R\left(\mathbf{\zeta}\right)$ shown in equation \ref{eq-app:0} where $\zeta$ is a point that lies on the line between $\mathbf{x}$ and reference point $\mathbf{a}$ and $\mathrm{\mathbf{H}}_{\mathbf{\zeta}}$ is the Hessian evaluated at point $\zeta$.
\begin{equation}
    %\begin{split}
    R\left(\mathbf{\zeta}\right) = \frac{1}{2}\left(\mathbf{x} - \mathbf{a}\right)^{T}\mathrm{\mathbf{H}}_{\mathbf{\zeta}}\left(\mathbf{x} - \mathbf{a}\right)
    %\mathrm{Tr} \left(\mathrm{H}_{\mathbf{\zeta}} \left(\mathbf{x} - \mathbf{a}\right)\left(\mathbf{x} - \mathbf{a}\right)^{T}\right)
    %\end{split}
    \label{eq-app:0}
\end{equation}
For active power flow from node $i$ to node $j$ this remainder term  takes the form shown in equation \ref{eq-app:1}.
\begin{equation}
    \begin{split}
        R\left(\zeta\right) =  
        g_{ij}
    \Delta_{v_{i}}^{2} + 
    |y_{ij}|\left[
    \frac{\zeta_{{v}_{i}}\zeta_{{v}_{j}}}{2}\cos\left(\zeta_{{\delta}_{ij}} - \angle _{y_{ij}} \right)\Delta_{\delta_{ij}}^{2} -
    \Lambda\sin\left(\zeta_{{\delta}_{ij}} - \angle _{y_{ij}} \right)
    \right]
     \\
    \Lambda = \zeta_{{v}_{i}}\Delta_{v_{j}}\Delta_{\delta_{ij}}  + \Delta_{v_{i}}\zeta_{{v}_{j}}\Delta_{\delta_{ij}}
    \end{split}
    \label{eq-app:1}
\end{equation}
An upper bound on the absolute value of the remainder can be formed by summing the absolute values of individual terms as seen in equation \ref{eq-app:2}
\begin{equation}
    \begin{split}
        \left|R\left(\zeta\right)\right| =  
        \left| g_{ij} \Delta_{v_{i}}^{2}  + 
    |y_{ij}|\left[
    \frac{\zeta_{{v}_{i}}\zeta_{{v}_{j}}}{2}\cos\left(\zeta_{{\delta}_{ij}} - \angle _{y_{ij}} \right)\Delta_{\delta_{ij}}^{2} -
    \Lambda\sin\left(\zeta_{{\delta}_{ij}} - \angle _{y_{ij}} \right)
    \right] \right|
     \\
     \leq \left| g_{ij} \Delta_{v_{i}}^{2} \right| + 
    \left|y_{ij}\frac{\zeta_{{v}_{i}}\zeta_{{v}_{j}}}{2}\cos\left(\zeta_{{\delta}_{ij}} - \angle _{y_{ij}} \right)\Delta_{\delta_{ij}}^{2}\right| +
    \left|y_{ij}\Lambda\sin\left(\zeta_{{\delta}_{ij}} - \angle _{y_{ij}} \right)\right|
    \\
    \leq \left| g_{ij} \Delta_{v_{i}}^{2} \right| + 
    \left|y_{ij}\frac{\zeta_{{v}_{i}}\zeta_{{v}_{j}}}{2}\Delta_{\delta_{ij}}^{2}\right| +
    \left|y_{ij}\Lambda\right|
    \\
    \leq \left| g_{ij} \Delta_{v_{i}}^{2} \right| + 
    \left|y_{ij}\right|\left[\left|\frac{{v}_{ub}^{2}}{2}\Delta_{\delta_{ij}}^{2}\right| +
    \left|{v}_{ub}\Delta_{v_{j}}\Delta_{\delta_{ij}}  \right|
    +
    \left| {v}_{ub}\Delta_{v_{i}}\Delta_{\delta_{ij}}\right|\right]
    \end{split}
    \label{eq-app:2}
\end{equation}
If we take the expectation of this upper bound on the remainder we get the first expression in equation \ref{eq-app:3}. If we assume that $v_{i}$, $v_{j}$ and $\delta_{ij}$ are independent and symmetric then this expectation is minimised by the mean values of $v_{i}$, $v_{j}$ and $\delta_{ij}$. If we do not want to make this assumption we can use the Cauchy-Schwartz inequality to find and upper bound on this expectation which is minimised by the mean.
\begin{equation}
    \begin{split}
        \mathbb{E}\left[|R\left(\zeta\right)|_{ub}\right] = \left| g_{ij}\right| \mathbb{E}\left[\Delta_{v_{i}}^{2} \right] + 
    {v}_{ub}\left|y_{ij}\right|\left[\frac{{v}_{ub}\mathbb{E}\left[\Delta_{\delta_{ij}}^{2}\right]}{2} +
    \mathbb{E}\left[\left|\Delta_{v_{j}}\Delta_{\delta_{ij}}  \right|\right]
    +
    \mathbb{E}\left[\left| \Delta_{v_{i}}\Delta_{\delta_{ij}}\right|\right]\right]
    \\
    \leq \left| g_{ij}\right| \mathbb{E}\left[\Delta_{v_{i}}^{2} \right] + 
    {v}_{ub}\left|y_{ij}\right|\left[\frac{{v}_{ub}\mathbb{E}\left[\Delta_{\delta_{ij}}^{2}\right]}{2} +
    \sqrt{\mathbb{E}\left[\Delta_{v_{j}}^{2}\right]\mathbb{E}\left[\Delta_{\delta_{ij}}^{2}\right]} +
    \sqrt{\mathbb{E}\left[\Delta_{v_{i}}^{2}\right]\mathbb{E}\left[\Delta_{\delta_{ij}}^{2}\right]}\right]
    \end{split}
    \label{eq-app:3}
\end{equation}
A similar process can be done for reactive power flow to derive the bound show in equation \ref{eq-app:4} as $b_{ij}$ is typically much larger $g_{ij}$ we can expect greater error in predicting reactive power flow than active power flow.
\begin{equation}
    \begin{split}
        \mathbb{E}\left[|R\left(\zeta\right)|_{ub}\right] = \left| b_{ij}\right| \mathbb{E}\left[\Delta_{v_{i}}^{2} \right] + 
    {v}_{ub}\left|y_{ij}\right|\left[\frac{{v}_{ub}\mathbb{E}\left[\Delta_{\delta_{ij}}^{2}\right]}{2} +
    \mathbb{E}\left[\left|\Delta_{v_{j}}\Delta_{\delta_{ij}}  \right|\right]
    +
    \mathbb{E}\left[\left| \Delta_{v_{i}}\Delta_{\delta_{ij}}\right|\right]\right]
    \\
    \leq \left| b_{ij}\right| \mathbb{E}\left[\Delta_{v_{i}}^{2} \right] + 
    {v}_{ub}\left|y_{ij}\right|\left[\frac{{v}_{ub}\mathbb{E}\left[\Delta_{\delta_{ij}}^{2}\right]}{2} +
    \sqrt{\mathbb{E}\left[\Delta_{v_{j}}^{2}\right]\mathbb{E}\left[\Delta_{\delta_{ij}}^{2}\right]} +
    \sqrt{\mathbb{E}\left[\Delta_{v_{i}}^{2}\right]\mathbb{E}\left[\Delta_{\delta_{ij}}^{2}\right]}\right]
    \end{split}
    \label{eq-app:4}
\end{equation}

\subsubsection{Data efficient approximation}

Consider the function $\mathbf{f}:\mathbf{R}^{n} \rightarrow \mathbf{R}^{1}$ with a Hessian $\mathbf{H}$ that is bounded by $\mathbf{M}$ so that the absolute values of the elements in $\mathbf{H}$ are less than the corresponding element in $\mathbf{M}$, $\left|\mathbf{H}\right| \leq \mathbf{M}$. If we examine its Taylor expansion as shown in equation \ref{eq-app:5} where the reference point is the mean of $\mathbf{x}$ we see that the absolute difference between the value of the function at the mean and the mean of the function value over $\mathbf{x}$ is expressed in terms of the hessian of the function and the covariance of $\mathbf{x}$. These equations extend into the multivariate case the work shown in \citet{3127971} (Licensed under CC BY-SA 3.0).

\begin{equation}
    \begin{split}
    \mathbf{f}(\mathbf{x}) = \mathbf{f}(\mathbf{\tilde{x}}) + 
    \mathbf{J}_{\mathbf{x}}\left(\mathbf{x} - \mathbf{\tilde{x}}\right) + 
    \frac{1}{2}\left(\mathbf{x} -\mathbf{\tilde{x}}\right)^{T}\mathrm{\mathbf{H}}_{\mathbf{\zeta}}\left(\mathbf{x} - \mathbf{\tilde{x}}\right) 
    \\
    \mathbf{f}(\mathbf{x}) - \mathbf{f}(\mathbf{\tilde{x}}) = 
    \mathbf{J}_{\mathbf{x}}\left(\mathbf{x} - \mathbf{\tilde{x}}\right) + 
    \frac{1}{2}\left(\mathbf{x} -\mathbf{\tilde{x}}\right)^{T}\mathrm{\mathbf{H}}_{\mathbf{\zeta}}\left(\mathbf{x} - \mathbf{\tilde{x}}\right)
    \\
    \mathbb{E}\left[\mathbf{f}(\mathbf{x})\right] - \mathbf{f}(\mathbf{\tilde{x}}) = 
    \mathbf{J}_{\mathbf{x}}\left(\mathbb{E}\left[\mathbf{x}\right] - \mathbf{\tilde{x}}\right) + 
    \frac{1}{2}%\left(\mathbf{x} -\mathbf{a}\right)^{T}\mathrm{\mathbf{H}}_{\mathbf{\zeta}}\left(\mathbf{x} - \mathbf{a}\right)
    \mathrm{Tr} \left(\mathrm{\mathbf{H}}_{\mathbf{\zeta}} \mathbb{E}\left[\left(\mathbf{x} - \mathbf{\tilde{x}}\right)\left(\mathbf{x} - \mathbf{\tilde{x}}\right)^{T}\right]\right) \\+ \frac{1}{2}\left(\mathbb{E}\left[\mathbf{x}\right] -\mathbf{\tilde{x}}\right)^{T}\mathrm{\mathbf{H}}_{\mathbf{\zeta}}\left(\mathbb{E}\left[\mathbf{x}\right] - \mathbf{\tilde{x}}\right) 
    \\
    \mathbb{E}\left[\mathbf{f}(\mathbf{x})\right] - \mathbf{f}\left(\mathbb{E}\left[\mathbf{x}\right]\right) = 
    \frac{1}{2}
    \mathrm{Tr} \left(\mathrm{\mathbf{H}}_{\mathbf{\zeta}} \Sigma_{\mathbf{x}} \right)
    \\
    \left|\mathbb{E}\left[\mathbf{f}(\mathbf{x})\right] - \mathbf{f}\left(\mathbb{E}\left[\mathbf{x}\right]\right)\right| = 
    \left|\frac{1}{2}
    \mathrm{Tr} \left(\mathrm{\mathbf{H}}_{\mathbf{\zeta}} \Sigma_{\mathbf{x}} \right)\right|
    \\
    \leq \frac{1}{2}
    \mathrm{Tr} \left(\left|\mathrm{\mathbf{H}}_{\mathbf{\zeta}}\right| \left|\Sigma_{\mathbf{x}}\right| \right)
    \\
    \leq \frac{1}{2}
    \mathrm{Tr} \left(\mathbf{M} \left|\Sigma_{\mathbf{x}}\right| \right)
    \end{split}
    \label{eq-app:5}
\end{equation}

Numerically, for the self-generated 118 node case we observed, the mean absolute difference over all nodes was $\mathbf{2.1217e-4}$, $\mathbf{4.0167e-4}$ for voltage magnitude and angle, respectively. Numerically, for the self-generated 30 node case we observed, the mean absolute difference over all nodes was $\mathbf{4.5786e-4}$, $\mathbf{9.8582e-4}$ for voltage magnitude and angle, respectively.

\subsubsection{Active power generation error comparison}
\begin{figure}[!htb]
\minipage{0.49\textwidth}
\includegraphics[width=\linewidth]{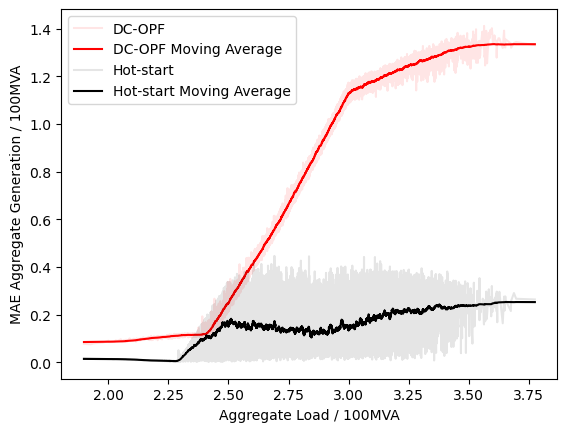}
\caption{Sum Absolute Error in in active power generation for all generators sorted by aggregate active power demand for the self-generated 30 node case dataset. This figure compares this error in DC-OPF and the hot-start linear power flow. This figure shows error in DC-OPF approximation is typically worse than for hot-start and that this error is dependent on aggregate demand and generally worsens as we increase aggregate demand, saturating at higher levels}
\endminipage\hfill
\minipage{0.49\textwidth}
\includegraphics[width=\linewidth]{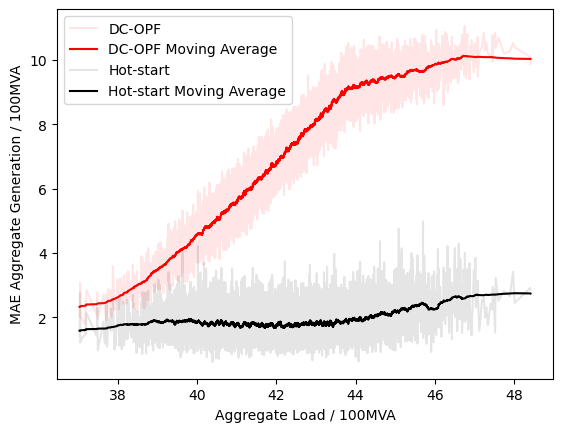}
\caption{Sum Absolute Error in in active power generation for all generators sorted by aggregate active power demand for the self-generated 118 node case dataset. This figure compares this error in DC-OPF and the hot-start linear power flow. This figure shows error in DC-OPF approximation is typically worse than for hot-start and that this error is dependent on aggregate demand and generally worsens as we increase aggregate demand, saturating at higher levels}
\endminipage
\end{figure}

\newpage
\subsection{Regression metrics}
The following tables show the regression metrics on the validation and train splits of the datasets considered.
\label{app:1}
\begin{table}[!htbp]
\caption{A table showing the MSE and FVU in predicting the voltage angle ($V_{a}$) in rad and voltage magnitude ($V_{m}$) in pu of the different methods considered for the train split on the self-generated dataset on the IEEE case 30 grid. There are 2 variations of the OPFormer-V shown, \textbf{feats-2} takes a 2 dimensional vector of load $(p_{l,i}, q_{l,i})$ as input while \textbf{feats-8} takes an eight dimensional vector of load, shunt susceptance and generator information $(p_{l,i}, q_{l,i}, bs_{i}, p^{max}_{g,i}, q^{max}_{g,i}, q^{min}_{g,i}, c1, c2)$. For NN methods we report the mean and standard deviation over 3 runs.}
\centering
\begin{tabular}{ccrr|rr}
\hline
\multicolumn{2}{c}{\multirow{2}{*}{\textbf{\begin{tabular}[c]{@{}c@{}}IEEE case30\\ (Train)\end{tabular}}}}                         & \multicolumn{2}{c|}{\textbf{$V_{a}$}}                                & \multicolumn{2}{c}{\textbf{$V_{m}$}}                                \\ \cline{3-6} 
\multicolumn{2}{c}{}                                                                                                                & \multicolumn{1}{c}{\textbf{MSE}} & \multicolumn{1}{c|}{\textbf{FVU}} & \multicolumn{1}{c}{\textbf{MSE}} & \multicolumn{1}{c}{\textbf{FVU}} \\ \hline
\multirow{2}{*}{\textbf{DeepOPF-V}}                                                      & \multicolumn{1}{c|}{\textbf{($\mu$)}}    & 8.746$\times 10^{-6}$                        & 2.275$\times 10^{-3}$                         & 3.260$\times 10^{-6}$                        & 9.128$\times 10^{-3}$                        \\
                                                                                         & \multicolumn{1}{c|}{\textbf{($\sigma$)}} & $\pm 5.3\times 10^{-7}$                          & $\pm 1.4\times 10^{-4}$                           & $\pm 1.3\times 10^{-7}$                          & $\pm 3.5\times 10^{-4}$                          \\
\multirow{2}{*}{\textbf{\begin{tabular}[c]{@{}c@{}}OPFormer-V\\ (feats 8)\end{tabular}}} & \multicolumn{1}{c|}{\textbf{($\mu$)}}    & 2.077$\times 10^{-7}$                        & 5.402$\times 10^{-5}$                         & 3.115$\times 10^{-8}$                        & 8.723$\times 10^{-5}$                        \\
                                                                                         & \multicolumn{1}{c|}{\textbf{($\sigma$)}} & $\pm 8.3\times 10^{-8}$                          & $\pm 2.2 \times 10^{-5}$                           & $\pm 4.3\times 10^{-9}$                          & $\pm 1.2\times 10^{-5}$                          \\
\multirow{2}{*}{\textbf{\begin{tabular}[c]{@{}c@{}}OPFormer-V\\ (feats 2)\end{tabular}}} & \multicolumn{1}{c|}{\textbf{($\mu$)}}    & 2.393$\times 10^{-7}$                        & 6.225$\times 10^{-5}$                         & 3.964$\times 10^{-8}$                        & 1.110$\times 10^{-4}$                        \\
                                                                                         & \multicolumn{1}{c|}{\textbf{($\sigma$)}} & $\pm 1.6\times 10^{-7}$                          & $\pm 4.1\times 10^{-5}$                           & $\pm 1.2\times 10^{-8}$                          & $\pm 3.5\times 10^{-5}$                          \\
\textbf{Grid Avg.}                                                                       & \multicolumn{1}{c|}{\textbf{}}           & 3.845$\times 10^{-3}$                        & 1.000$\times 10^{-0}$                         & 3.571$\times 10^{-4}$                        & 1.000$\times 10^{-0}$                        \\
\textbf{Node Avg.}                                                                       & \multicolumn{1}{c|}{\textbf{}}           & 9.768$\times 10^{-5}$                        & 2.540$\times 10^{-2}$                         & 2.726$\times 10^{-5}$                        & 7.633$\times 10^{-2}$                        \\
\textbf{DC-OPF}                                                                           & \multicolumn{1}{c|}{\textbf{}}           & 1.925$\times 10^{-3}$                        & 5.007$\times 10^{-1}$                         & 9.490$\times 10^{-4}$                        & 2.657$\times 10^{-0}$                        \\
\textbf{Linear}                                                                          & \multicolumn{1}{c|}{\textbf{}}           & 6.328$\times 10^{-6}$                        & 1.646$\times 10^{-3}$                         & 4.464$\times 10^{-7}$                        & 1.250$\times 10^{-3}$                        \\
\textbf{GP}                                                                              & \multicolumn{1}{c|}{\textbf{}}           & 1.225$\times 10^{-17}$                        & 3.186$\times 10^{-15}$                         & 4.082$\times 10^{-11}$                        & 1.143$\times 10^{-7}$                        \\
\textbf{Hot-Start}                                                                              & \multicolumn{1}{c|}{\textbf{}}           & 1.899$\times 10^{-5}$                       & 4.939$\times 10^{-3}$                         & 4.484$\times 10^{-5}$                         & 1.256$\times 10^{-1}$                        \\ \hline
\end{tabular}
\end{table}

\begin{table}[!htbp]
\caption{A table showing the MSE and FVU in predicting the voltage angle ($V_{a}$) in rad and voltage magnitude ($V_{m}$) in pu of the different methods considered for the validation split on the self-generated dataset on the IEEE case 30 grid. There are 2 variations of the OPFormer-V shown, \textbf{feats-2} takes a 2 dimensional vector of load $(p_{l,i}, q_{l,i})$ as input while \textbf{feats-8} takes an eight dimensional vector of load, shunt susceptance and generator information $(p_{l,i}, q_{l,i}, bs_{i}, p^{max}_{g,i}, q^{max}_{g,i}, q^{min}_{g,i}, c1, c2)$. For NN methods we report the mean and standard deviation over 3 runs.}
\centering
\begin{tabular}{ccrr|rr}
\hline
\multicolumn{2}{c}{\multirow{2}{*}{\textbf{\begin{tabular}[c]{@{}c@{}}IEEE case30\\ (Val.)\end{tabular}}}}                          & \multicolumn{2}{c|}{\textbf{$V_{a}$}}                                & \multicolumn{2}{c}{\textbf{$V_{m}$}}                                \\ \cline{3-6} 
\multicolumn{2}{c}{}                                                                                                                & \multicolumn{1}{c}{\textbf{MSE}} & \multicolumn{1}{c|}{\textbf{FVU}} & \multicolumn{1}{c}{\textbf{MSE}} & \multicolumn{1}{c}{\textbf{FVU}} \\ \hline
\multirow{2}{*}{\textbf{DeepOPF-V}}                                                      & \multicolumn{1}{c|}{\textbf{($\mu$)}}    & 8.778$\times 10^{-6}$                        & 2.282$\times 10^{-3}$                         & 3.281$\times 10^{-6}$                        & 9.204$\times 10^{-3}$                        \\
                                                                                         & \multicolumn{1}{c|}{\textbf{($\sigma$)}} & $\pm 5.3\times 10^{-7}$                          & $\pm 1.4\times 10^{-4}$                           & $\pm 1.2\times 10^{-7}$                          & $\pm 3.5\times 10^{-4}$                          \\
\multirow{2}{*}{\textbf{\begin{tabular}[c]{@{}c@{}}OPFormer-V\\ (feats 8)\end{tabular}}} & \multicolumn{1}{c|}{\textbf{($\mu$)}}    & 2.139$\times 10^{-7}$                        & 5.560$\times 10^{-5}$                         & 3.208$\times 10^{-8}$                        & 9.001$\times 10^{-5}$                        \\
                                                                                         & \multicolumn{1}{c|}{\textbf{($\sigma$)}} & $\pm 8.5\times 10^{-8}$                          & $\pm 2.2\times 10^{-5}$                           & $\pm 4.4\times 10^{-9}$                          & $\pm 1.2\times 10^{-5}$                          \\
\multirow{2}{*}{\textbf{\begin{tabular}[c]{@{}c@{}}OPFormer-V\\ (feats 2)\end{tabular}}} & \multicolumn{1}{c|}{\textbf{($\mu$)}}    & 2.413$\times 10^{-7}$                        & 6.272$\times 10^{-5}$                         & 4.036$\times 10^{-8}$                        & 1.132$\times 10^{-4}$                        \\
                                                                                         & \multicolumn{1}{c|}{\textbf{($\sigma$)}} & $\pm 1.6\times 10^{-7}$                          & $\pm 4.0\times 10^{-5}$                           & $\pm 1.3 \times 10^{-8}$                          & $\pm 3.6\times 10^{-5}$                          \\
\textbf{Grid Avg.}                                                                       & \multicolumn{1}{c|}{\textbf{}}           & 3.847$\times 10^{-3}$                        & 1.000$\times 10^{-0}$                         & 3.564$\times 10^{-4}$                        & 1.000$\times 10^{-0}$                        \\
\textbf{Node Avg.}                                                                       & \multicolumn{1}{c|}{\textbf{}}           & 9.799$\times 10^{-5}$                        & 2.547$\times 10^{-2}$                         & 2.735$\times 10^{-5}$                        & 7.674$\times 10^{-2}$                        \\
\textbf{DC-OPF}                                                                           & \multicolumn{1}{c|}{\textbf{}}           & 1.915$\times 10^{-3}$                        & 4.977$\times 10^{-1}$                         & 9.504$\times 10^{-4}$                        & 2.666$\times 10^{-0}$                        \\
\textbf{Linear}                                                                          & \multicolumn{1}{c|}{\textbf{}}           & 6.538$\times 10^{-6}$                        & 1.700$\times 10^{-3}$                         & 4.604$\times 10^{-7}$                        & 1.292$\times 10^{-3}$                        \\
\textbf{GP}                                                                              & \multicolumn{1}{c|}{\textbf{}}           & 1.200$\times 10^{-6}$                        & 3.118$\times 10^{-4}$                         & 1.002$\times 10^{-7}$                        & 2.811$\times 10^{-4}$                        \\ 
\textbf{Hot-Start}                                                                              & \multicolumn{1}{c|}{\textbf{}}           & 1.900$\times 10^{-5}$                        & 4.939$\times 10^{-3}$                         & 4.446$\times 10^{-5}$                        & 1.247$\times 10^{-1}$                        \\ \hline
\end{tabular}
\end{table}

\begin{table}[!htbp]
\caption{A table showing the MSE and FVU in predicting the voltage angle ($V_{a}$) in rad and voltage magnitude ($V_{m}$) in pu of the different methods considered for the train split on the self-generated dataset on the IEEE case 118 grid. The OPFormer-V variation considered \textbf{feats-8} takes an eight dimensional vector of load, shunt susceptance and generator information $(p_{l,i}, q_{l,i}, bs_{i}, p^{max}_{g,i}, q^{max}_{g,i}, q^{min}_{g,i}, c1, c2)$. For NN methods we report the mean and standard deviation over 3 runs.}
\centering
\begin{tabular}{ccrr|rr}
\hline
\multicolumn{2}{c}{\multirow{2}{*}{\textbf{\begin{tabular}[c]{@{}c@{}}IEEE case118\\ (Train)\end{tabular}}}} & \multicolumn{2}{c|}{\textbf{$V_{a}$}}                                & \multicolumn{2}{c}{\textbf{$V_{m}$}}                                \\ \cline{3-6} 
\multicolumn{2}{c}{}                                                                                         & \multicolumn{1}{c}{\textbf{MSE}} & \multicolumn{1}{c|}{\textbf{FVU}} & \multicolumn{1}{c}{\textbf{MSE}} & \multicolumn{1}{c}{\textbf{FVU}} \\ \hline
\multirow{2}{*}{\textbf{DeepOPF-V}}                 & \multicolumn{1}{c|}{\textbf{($\mu$)}}                  & 7.374$\times 10^{-5}$                        & 1.093$\times 10^{-2}$                         & 6.473$\times 10^{-6}$                        & 4.140$\times 10^{-2}$                        \\
                                                    & \multicolumn{1}{c|}{\textbf{($\sigma$)}}               & $\pm 4.3\times 10^{-6}$                          & $\pm 6.4\times 10^{-4}$                           & $\pm 9.7\times 10^{-8}$                          & $\pm 6.2\times 10^{-4}$                          \\
\multirow{2}{*}{\textbf{\begin{tabular}[c]{@{}c@{}}OPFormer-V\\ (feats 8)\end{tabular}}}                & \multicolumn{1}{c|}{\textbf{($\mu$)}}                  & 2.661$\times 10^{-6}$                        & 3.943$\times 10^{-4}$                         & 6.570$\times 10^{-8}$                        & 4.202$\times 10^{-4}$                        \\
                                                    & \multicolumn{1}{c|}{\textbf{($\sigma$)}}               & $\pm 3.1 \times 10^{-7}$                          & $\pm 4.7 \times 10^{-5}$                           & $\pm 5.4 \times 10^{-9}$                          & $\pm 3.4 \times 10^{-5}$                          \\
\multicolumn{2}{c|}{\textbf{Grid Avg.}}                                                                      & 6.749$\times 10^{-3}$                        & 1.000$\times 10^{-0}$                         & 1.563$\times 10^{-4}$                        & 1.000$\times 10^{-0}$                        \\
\multicolumn{2}{c|}{\textbf{Node Avg.}}                                                                      & 4.428$\times 10^{-4}$                        & 6.560$\times 10^{-2}$                         & 7.550$\times 10^{-6}$                        & 4.829$\times 10^{-2}$                        \\
\multicolumn{2}{c|}{\textbf{DC-OPF}}                                                                          & 4.573$\times 10^{-3}$                        & 6.775$\times 10^{-1}$                         & 1.795$\times 10^{-3}$                        & 1.148$\times 10^{+1}$                        \\
\multicolumn{2}{c|}{\textbf{Linear}}                                                                         & 2.948$\times 10^{-6}$                        & 4.368$\times 10^{-4}$                         & 1.178$\times 10^{-7}$                        & 7.534$\times 10^{-4}$                        \\
\multicolumn{2}{c|}{\textbf{GP}}                                                                             & 6.290$\times 10^{-6}$                        & 9.320$\times 10^{-4}$                         & 2.595$\times 10^{-7}$                        & 1.660$\times 10^{-3}$                        \\
\textbf{Hot-Start}                                                                              & \multicolumn{1}{c|}{\textbf{}}           & 1.172$\times 10^{-3}$                        & 1.737$\times 10^{-1}$                         & 2.526$\times 10^{-3}$                        & 1.616$\times 10^{+1}$                        \\ \hline
\end{tabular}
\end{table}

\begin{table}[!htbp]
\caption{A table showing the MSE and FVU in predicting the voltage angle ($V_{a}$) in rad and voltage magnitude ($V_{m}$) in pu of the different methods considered for the validation split on the self-generated dataset on the IEEE case 118 grid. The OPFormer-V variation considered \textbf{feats-8} takes an eight dimensional vector of load, shunt susceptance and generator information $(p_{l,i}, q_{l,i}, bs_{i}, p^{max}_{g,i}, q^{max}_{g,i}, q^{min}_{g,i}, c1, c2)$. For NN methods we report the mean and standard deviation over 3 runs.}
\centering
\begin{tabular}{ccrr|rr}
\hline
\multicolumn{2}{c}{\multirow{2}{*}{\textbf{\begin{tabular}[c]{@{}c@{}}IEEE case118\\ (Val.)\end{tabular}}}}                         & \multicolumn{2}{c|}{\textbf{$V_{a}$}}                                & \multicolumn{2}{c}{\textbf{$V_{m}$}}                                \\ \cline{3-6} 
\multicolumn{2}{c}{}                                                                                                                & \multicolumn{1}{c}{\textbf{MSE}} & \multicolumn{1}{c|}{\textbf{FVU}} & \multicolumn{1}{c}{\textbf{MSE}} & \multicolumn{1}{c}{\textbf{FVU}} \\ \hline
\multirow{2}{*}{\textbf{DeepOPF-V}}                                                      & \multicolumn{1}{c|}{\textbf{($\mu$)}}    & 7.326$\times 10^{-5}$                        & 1.085$\times 10^{-2}$                         & 6.465$\times 10^{-6}$                        & 4.133$\times 10^{-2}$                        \\
                                                                                         & \multicolumn{1}{c|}{\textbf{($\sigma$)}} & $\pm 4.1 \times 10^{-6}$                          & $\pm 6.1 \times 10^{-4}$                           & $\pm 9.4 \times 10^{-8}$                          & $\pm 6.0 \times 10^{-4}$                          \\
\multirow{2}{*}{\textbf{\begin{tabular}[c]{@{}c@{}}OPFormer-V\\ (feats 8)\end{tabular}}} & \multicolumn{1}{c|}{\textbf{($\mu$)}}    & 2.774$\times 10^{-6}$                        & 4.110$\times 10^{-4}$                         & 6.755$\times 10^{-8}$                        & 4.319$\times 10^{-4}$                        \\
                                                                                         & \multicolumn{1}{c|}{\textbf{($\sigma$)}} & $\pm 3.1 \times 10^{-7}$                          & $\pm 4.5 \times 10^{-5}$                           & $\pm 5.7 \times 10^{-9}$                          & $\pm 3.6 \times 10^{-5}$                          \\
\textbf{Grid Avg.}                                                                       & \multicolumn{1}{c|}{\textbf{}}           & 6.750$\times 10^{-3}$                        & 1.000$\times 10^{-0}$                         & 1.564$\times 10^{-4}$                        & 1.000$\times 10^{-0}$                        \\
\textbf{Node Avg.}                                                                       & \multicolumn{1}{c|}{\textbf{}}           & 4.445$\times 10^{-4}$                        & 6.585$\times 10^{-2}$                         & 7.541$\times 10^{-6}$                        & 4.821$\times 10^{-2}$                        \\
\textbf{DC-OPF}                                                                           & \multicolumn{1}{c|}{\textbf{}}           & 4.540$\times 10^{-3}$                        & 6.726$\times 10^{-1}$                         & 1.795$\times 10^{-3}$                        & 1.147$\times 10^{+1}$                        \\
\textbf{Linear}                                                                          & \multicolumn{1}{c|}{\textbf{}}           & 3.049$\times 10^{-6}$                        & 4.516$\times 10^{-4}$                         & 1.191$\times 10^{-7}$                        & 7.612$\times 10^{-4}$                        \\
\textbf{GP}                                                                              & \multicolumn{1}{c|}{\textbf{}}           & 8.556$\times 10^{-6}$                        & 1.267$\times 10^{-3}$                         & 3.804$\times 10^{-7}$                        & 2.432$\times 10^{-3}$                        \\
\textbf{Hot-Start}                                                                              & \multicolumn{1}{c|}{\textbf{}}           & 1.196$\times 10^{-3}$                        & 1.772$\times 10^{-1}$                         & 2.526$\times 10^{-3}$                        & 1.615$\times 10^{+1}$                        \\ \hline
\end{tabular}
\end{table}

\begin{table}[!htbp]
\caption{A table showing the MSE and FVU in predicting the voltage angle ($V_{a}$) in rad and voltage magnitude ($V_{m}$) in pu for the different methods considered on the train split on the OPF-Learn case 30 dataset. The OPFormer-V variation considered \textbf{feats-2} takes a 2 dimensional vector of load $(p_{l,i}, q_{l,i})$ as input. For NN methods we report the mean and standard deviation over 3 runs.}
\centering
\begin{tabular}{ccrr|rr}
\hline
\multicolumn{2}{c}{\multirow{2}{*}{\textbf{\begin{tabular}[c]{@{}c@{}}OPF-Learn case30\\ (Train)\end{tabular}}}}                    & \multicolumn{2}{c|}{\textbf{$V_{a}$}}                                & \multicolumn{2}{c}{\textbf{$V_{m}$}}                                \\ \cline{3-6} 
\multicolumn{2}{c}{}                                                                                                                & \multicolumn{1}{c}{\textbf{MSE}} & \multicolumn{1}{c|}{\textbf{FVU}} & \multicolumn{1}{c}{\textbf{MSE}} & \multicolumn{1}{c}{\textbf{FVU}} \\ \hline
\multirow{2}{*}{\textbf{DeepOPF-V}}                                                      & \multicolumn{1}{c|}{\textbf{($\mu$)}}    & 4.227$\times 10^{-7}$                        & 1.368$\times 10^{-1}$                         & 2.384$\times 10^{-4}$                        & 2.762$\times 10^{-1}$                        \\
                                                                                         & \multicolumn{1}{c|}{\textbf{($\sigma$)}} & $\pm 2.3 \times 10^{-12}$                          & $\pm 7.4 \times 10^{-7}$                           & $\pm 2.0\times 10^{-9}$                          & $\pm 2.3\times 10^{-6}$                          \\
\multirow{2}{*}{\textbf{\begin{tabular}[c]{@{}c@{}}OPFormer-V\\ (feats 2)\end{tabular}}} & \multicolumn{1}{c|}{\textbf{($\mu$)}}    & 3.388$\times 10^{-7}$                        & 1.096$\times 10^{-1}$                         & 2.194$\times 10^{-4}$                        & 2.541$\times 10^{-1}$                        \\
                                                                                         & \multicolumn{1}{c|}{\textbf{($\sigma$)}} & $\pm 7.2 \times 10^{-8}$                          & $\pm 2.3 \times 10^{-2}$                           & $\pm 1.7 \times 10^{-5}$                          & $\pm 2.0\times 10^{-2}$                          \\
\textbf{Grid Avg.}                                                                       & \multicolumn{1}{c|}{\textbf{}}           & 3.091$\times 10^{-6}$                        & 1.000$\times 10^{-0}$                         & 8.632$\times 10^{-4}$                        & 1.000$\times 10^{-0}$                        \\
\textbf{Node Avg.}                                                                       & \multicolumn{1}{c|}{\textbf{}}           & 4.227$\times 10^{-7}$                        & 1.368$\times 10^{-1}$                         & 2.384$\times 10^{-4}$                        & 2.762$\times 10^{-1}$                        \\
\textbf{Linear}                                                                          & \multicolumn{1}{c|}{\textbf{}}           & 6.426$\times 10^{-9}$                        & 2.079$\times 10^{-3}$                         & 3.350$\times 10^{-5}$                        & 3.881$\times 10^{-2}$                        \\
\textbf{GP}                                                                              & \multicolumn{1}{c|}{\textbf{}}           & 5.099$\times 10^{-9}$                        & 1.650$\times 10^{-3}$                         & 2.170$\times 10^{-5}$                        & 2.514$\times 10^{-2}$                        \\ \hline
\end{tabular}
\end{table}

\begin{table}[!htbp]
\caption{A table showing the MSE and FVU in predicting the voltage angle ($V_{a}$) in rad and voltage magnitude ($V_{m}$) in pu for the different methods considered on the validation split on the OPF-Learn case 30 dataset. The OPFormer-V variation considered \textbf{feats-2} takes a 2 dimensional vector of load $(p_{l,i}, q_{l,i})$ as input. For NN methods we report the mean and standard deviation over 3 runs.}
\centering
\begin{tabular}{ccrr|rr}
\hline
\multicolumn{2}{c}{\multirow{2}{*}{\textbf{\begin{tabular}[c]{@{}c@{}}OPF-Learn case30\\ (Val.)\end{tabular}}}}                     & \multicolumn{2}{c|}{\textbf{$V_{a}$}}                                & \multicolumn{2}{c}{\textbf{$V_{m}$}}                                \\ \cline{3-6} 
\multicolumn{2}{c}{}                                                                                                                & \multicolumn{1}{c}{\textbf{MSE}} & \multicolumn{1}{c|}{\textbf{FVU}} & \multicolumn{1}{c}{\textbf{MSE}} & \multicolumn{1}{c}{\textbf{FVU}} \\ \hline
\multirow{2}{*}{\textbf{DeepOPF-V}}                                                      & \multicolumn{1}{c|}{\textbf{($\mu$)}}    & 4.278$\times 10^{-7}$                        & 1.374$\times 10^{-1}$                         & 2.468$\times 10^{-4}$                        & 2.831$\times 10^{-1}$                        \\
                                                                                         & \multicolumn{1}{c|}{\textbf{($\sigma$)}} & $\pm 4.0\times 10^{-12}$                          & $\pm 1.3\times 10^{-6}$                           & $\pm 4.3\times 10^{-9}$                          & $\pm 4.9\times 10^{-6}$                          \\
\multirow{2}{*}{\textbf{\begin{tabular}[c]{@{}c@{}}OPFormer-V\\ (feats 2)\end{tabular}}} & \multicolumn{1}{c|}{\textbf{($\mu$)}}    & 3.412$\times 10^{-7}$                        & 1.096$\times 10^{-1}$                         & 2.268$\times 10^{-4}$                        & 2.602$\times 10^{-1}$                        \\
                                                                                         & \multicolumn{1}{c|}{\textbf{($\sigma$)}} & $\pm 7.4\times 10^{-8}$                          & $\pm 2.4\times 10^{-2}$                           & $\pm 1.8 \times 10^{-5}$                          & $\pm 2.1 \times 10^{-2}$                          \\
\textbf{Grid Avg.}                                                                       & \multicolumn{1}{c|}{\textbf{}}           & 3.115$\times 10^{-6}$                        & 1.000$\times 10^{-0}$                         & 8.715$\times 10^{-4}$                        & 1.000$\times 10^{-0}$                        \\
\textbf{Node Avg.}                                                                       & \multicolumn{1}{c|}{\textbf{}}           & 4.278$\times 10^{-7}$                        & 1.374$\times 10^{-1}$                         & 2.468$\times 10^{-4}$                        & 2.831$\times 10^{-1}$                        \\
\textbf{Linear}                                                                          & \multicolumn{1}{c|}{\textbf{}}           & 6.617$\times 10^{-9}$                        & 2.124$\times 10^{-3}$                         & 3.336$\times 10^{-5}$                        & 3.828$\times 10^{-2}$                        \\
\textbf{GP}                                                                              & \multicolumn{1}{c|}{\textbf{}}           & 6.680$\times 10^{-9}$                        & 2.145$\times 10^{-3}$                         & 3.431$\times 10^{-5}$                        & 3.937$\times 10^{-2}$                        \\ \hline
\end{tabular}
\end{table}

\newpage
\subsection{Power metrics}
The following tables show the power metrics on the test splits of the datasets considered for the non-ML approaches. For linear power flow methods, full AC-OPF solutions were generated using predicted voltages in the power flow equations, hence the violation in generation. An alternative approach could use predicted generator output and automatically satisfy generation constraints, but result in potential voltage violations.
\label{app:2}
\begin{table}[!htbp]
\caption{A table comparing the quality of the OPF solutions from the predictions of the other methods considered on the test split on the IEEE case30 datasets. Predictions are assessed on the relative gap from optimality, the rate of violation of generation limits, the relative difference between load in the ground truth and effective load derived using predicted voltage for both a grid aggregation and at a nodal level for $\neq 0$ loads.}
\label{tab:ieee30_power_appendix}
\centering
\begin{tabular}{l|r|r|r|r|r|r}
\hline
\multicolumn{1}{c|}{\textbf{IEEE case30}}             & \multicolumn{1}{c|}{\textbf{Grid}} & \multicolumn{1}{c|}{\textbf{Node}} & \multicolumn{1}{c|}{\textbf{DC-OPF}} & \multicolumn{1}{c|}{\textbf{OLS}} & \multicolumn{1}{c|}{\textbf{GP}} & \multicolumn{1}{c}{\textbf{Hot-Start}} \\ \hline
\textbf{Rel. Opt. Diff. (\%)}                    & 37.373                                  & -0.327                                  & 6.383                               & 0.005                                & -0.002 & -0.033                         \\
\textbf{Abs. Rel. Opt. Diff. (\%)}               & 37.373                                  & 4.271                                   & 6.383                               & 0.029                                & 0.087 & 0.143                          \\
\textbf{$\mathbf{P_{g}}$ Violation Rate (\%)}         & 23.967                                  & 4.467                                   & 2.097                               & 9.097                                & 11.323 & 7.796                         \\
\textbf{$\mathbf{Q_{g}}$ Violation Rate (\%)}         & 17.949                                  & 24.149                                  & 32.979                              & 15.222                               & 13.606 & 28.670                         \\
\textbf{Abs. Rel. Tot. $\mathbf{P_{d}}$ (\%)}    & 48.987                                  & 3.284                                   & 5.931                               & 0.018                                & 0.066 & 0.122                          \\
\textbf{Abs. Rel. Tot. $\mathbf{Q_{d}}$ (\%)}    & 22.308                                  & 3.892                                   & 73.524                              & 0.088                                & 0.067 & 0.912                           \\
\textbf{Abs. Rel. $\mathbf{P^{\neq 0}_{d}}$ (\%)} & 85.714                                  & 24.660                                  & 29.461                              & 0.197                                & 0.242 & 0.258                          \\
\textbf{Abs. Rel. $\mathbf{Q^{\neq 0}_{d}}$ (\%)} & 248.173                                 & 24.760                                  & 431.436                             & 0.326                                & 0.309 & 1.733                          \\ \hline
\end{tabular}
\end{table}
\begin{table}[!htbp]
\caption{A table comparing the quality of the OPF solutions from the predictions of the other methods considered on the test split on the IEEE case118 datasets. Predictions are assessed on the relative gap from optimality, the rate of violation of generation limits, the relative difference between load in the ground truth and effective load derived using predicted voltage for both a grid aggregation and at a nodal level for $\neq 0$ loads.}
\label{tab:ieee118_power_appendix}
\centering
\begin{tabular}{l|r|r|r|r|r|r}
\hline
\multicolumn{1}{c|}{\textbf{IEEE case118}}             & \multicolumn{1}{c|}{\textbf{Grid}} & \multicolumn{1}{c|}{\textbf{Node}} & \multicolumn{1}{c|}{\textbf{DC-OPF}} & \multicolumn{1}{c|}{\textbf{OLS}} & \multicolumn{1}{c}{\textbf{GP}} & \multicolumn{1}{c}{\textbf{Hot-Start}} \\ \hline
\textbf{Rel. Opt. Diff. (\%)}                    & 14.904                                  & -0.831                                  & 2.057                               & 0.002                                & -0.107   & -0.378                       \\
\textbf{Abs. Rel. Opt. Diff. (\%)}               & 14.904                                  & 1.772                                   & 2.057                               & 0.012                                & 2.212   & 0.382                        \\
\textbf{$\mathbf{P_{g}}$ Violation Rate (\%)}         & 16.953                                  & 21.645                                  & 4.221                               & 15.201                               & 16.509   & 14.025                       \\
\textbf{$\mathbf{Q_{g}}$ Violation Rate (\%)}         & 14.437                                  & 13.263                                  & 27.334                              & 10.691                               & 11.914   & 28.673                       \\
\textbf{Abs. Rel. Tot. $\mathbf{P_{d}}$ (\%)}    & 33.798                                  & 1.254                                   & 2.186                               & 0.009                                & 1.673   & 0.381                        \\
\textbf{Abs. Rel. Tot. $\mathbf{Q_{d}}$ (\%)}    & 26.182                                  & 1.472                                   & 60.329                              & 0.088                                & 1.715   & 1.548                        \\
\textbf{Abs. Rel. $\mathbf{P^{\neq 0}_{d}}$ (\%)} & 54.545                                  & 15.707                                  & 6.520                               & 0.100                                & 3.171     & 1.383                      \\
\textbf{Abs. Rel. $\mathbf{Q^{\neq 0}_{d}}$ (\%)} & 90.601                                  & 17.028                                  & 149.715                             & 0.467                                & 5.175    & 4.328                       \\ \hline
\end{tabular}
\end{table}

\newpage
\subsection{Additional experiment information}
\label{app:3}
\subsubsection{Transformer architecture overview}
%Transformer architecture overview
\begin{itemize}
    \item \textbf{input size}: 8 or 2
    \item \textbf{num. layers}: 7
    \item \textbf{num. transformer encoder layers}: 4
    \item \textbf{dim. ff}: 512
    \item \textbf{num. attn. heads}: 4
    \item \textbf{c hidden}: 16
    \item \textbf{c out}: 2 * (num. nodes) 
    \item \textbf{dropout rate}: 0.1
    \item \textbf{num. parameters:} $\sim$104k (30 nodes), 574k (118 nodes)
\end{itemize}

\subsubsection{MLP architecture overview}
\begin{itemize}
    \item \textbf{input size}: 2 * (num. nonzero load nodes) 
    \item \textbf{num. layers}: 8
    \item \textbf{c hidden}: 256 (for 30 nodes), 1024 (for 118 nodes)
    \item \textbf{c out}: 2 * (num. nodes) 
    \item \textbf{dropout rate}: 0.1
    \item \textbf{num. parameters:} $\sim$359k (30 nodes), 5.7M (118 nodes)
\end{itemize}

\subsubsection{Optimizer details}
\begin{itemize}
    \item \textbf{optimizer}: SGD 
    \item \textbf{learning rate}: 1e-3
    \item \textbf{weight decay}: 2e-6
    \item \textbf{momentum}: 0.9
    \item \textbf{lr scheduler}: Cosine Annealing
    \item \textbf{num. epochs}: 200
\end{itemize}

\subsubsection{Compute resources \& approximate run times}
Models trained on CPU (Apple M1 Pro Chip). For the 30 node case approximate train time of 2 hours and the 118 node case approximate train time of 5 hours. OPFLearn case30 approximate train time of 0.5 hours.

\newpage
\subsubsection{Speed-ups, MAC \& approximate parameter count}
In Tables~\ref{tab:mac0} and \ref{tab:mac1}, we report speed-ups, multiply–accumulate operations (MAC), and parameter counts observed in our experiments. Note that the AC-OPF solver was run on online resources (MATLAB Online), while ML models were trained on a CPU (Apple M1 Pro, 16GB RAM).  

\begin{table}[!htbp]
\centering
\begin{tabular}{lcc}
\toprule
\textbf{Metric} & \textbf{DeepOPF-V} & \textbf{OPFormer-V (feats 2)} \\
\midrule
MAC              & 5.351M   & 441K \\
Parameter Count  & 5.357M   & 43.2K \\
Approx. Speedup  & $\times$446 & $\times$717 \\
\bottomrule
\end{tabular}
\caption{Estimated MAC, parameter count, and speedup of DeepOPF-V and OPFormer-V (feats 2) on the 30-node case.}
\label{tab:mac0}
\end{table}

\begin{table}[!htbp]
\centering
\begin{tabular}{lcc}
\toprule
\textbf{Metric} & \textbf{DeepOPF-V} & \textbf{OPFormer-V (feats 2)} \\
\midrule
MAC              & 5.743M   & 2.891M \\
Parameter Count  & 5.750M   & 449K \\
Approx. Speedup  & $\times$553 & $\times$257 \\
\bottomrule
\end{tabular}
\caption{Estimated MAC, parameter count, and speedup of DeepOPF-V and OPFormer-V (feats 2) on the 118-node case.}
\label{tab:mac1}
\end{table}

\end{document}